\documentclass[twocolumn,preprintnumbers,amsmath,amssymb,superscriptaddress]{revtex4}
\usepackage{graphicx}
\usepackage{dcolumn}
\usepackage{bm}

\begin{document}
\title{Lead monoxide $\alpha$-PbO: electronic properties and point defect formation.}
\author{J. Berashevich $^\ddag$}
\affiliation{Thunder Bay Regional Research Institute, 290 Munro St., Thunder Bay, ON, P7B 5E1, Canada}
\author{O. Semeniuk}
\affiliation{Thunder Bay Regional Research Institute, 290 Munro St., Thunder Bay, ON, P7B 5E1, Canada}
\affiliation{Department of Physics, Lakehead University, 955 Oliver Road, Thunder Bay, ON, P7B 5E1}
\author{O. Rubel}
\affiliation{Thunder Bay Regional Research Institute, 290 Munro St., Thunder Bay, ON, P7B 5E1, Canada}
\affiliation{Department of Physics, Lakehead University, 955 Oliver Road, Thunder Bay, ON, P7B 5E1}
\author{J.A. Rowlands}
\affiliation{Thunder Bay Regional Research Institute, 290 Munro St., Thunder Bay, ON, P7B 5E1, Canada}
\author{A. Reznik}
\affiliation{Thunder Bay Regional Research Institute, 290 Munro St., Thunder Bay, ON, P7B 5E1, Canada}
\affiliation{Department of Physics, Lakehead University, 955 Oliver Road, Thunder Bay, ON, P7B 5E1}

\begin{abstract}
The electronic properties of polycrystalline lead oxide
consisting of a network of single-crystalline $\alpha$-PbO platelets and the
formation of the native point defects in $\alpha$-PbO crystal lattice are studied 
using first principles calculations.
The $\alpha$-PbO lattice consists of coupled layers interaction between which is 
too low to produce high efficiency interlayer charge transfer. 
In practice, the polycrystalline nature of $\alpha$-PbO causes the formation 
of lattice defects in such a high concentration that defect-related conductivity becomes the dominant factor 
in the interlayer charge transition. 
We found that the formation energy for the O vacancies 
is low, such vacancies are occupied by two electrons in the zero charge state and
tend to initiate the ionization interactions with the Pb vacancies.
The vacancies introduce localized states in the band gap which can affect charge transport. 
The O vacancy forms a defect state at 1.03 eV above the valence band which can act as 
a deep trap for electrons,
while the Pb vacancy forms a shallow trap for holes located just 0.1 eV above the valence band.
Charge de-trapping from O vacancies can be accounted for the experimentally found dark current decay 
in ITO/PbO/Au structures.
\end{abstract}

\maketitle
\section{Introduction}
Polycrystalline lead oxide (PbO) is one of the most promising photoconductive 
materials for use as a x-ray-to-charge transducer in direct conversion x-ray detectors \cite{simon}. 
Since the direct conversion detection scheme offers a number of advantages over the 
indirect conversion, \cite{Sasha}, photoconductive materials for x-ray 
imaging have recently attracted much interest.
The four most important criteria when potential x-ray photoconductors are considered include: 
(1) high conversion gain; (2) high x-ray absorption efficiency; 
(3) compatibility with large area detector technology and 
(4) good photoconductive properties. 
PbO satisfies the first three criteria. However, thick PbO does not show adequate
transport properties and this results in poor temporal characteristics (signal propagation/delay time) 
\cite{simon,hughes}. This is the primary issue in the development of 
PbO in x-ray medical imaging detectors.

\begin{figure}
\includegraphics[scale=0.8]{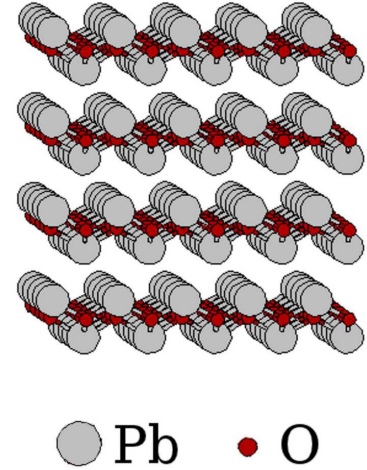}
\caption{\label{fig:fig1} The crystal structure of the 
tetragonal $\alpha$-PbO of the space group 129P4/nmm.}
\end{figure}

For use in x-ray detectors, photoconductive layers are deposited directly over an imaging matrix.  
When PbO layer is deposited in the evaporation process, it condenses in
very thin platelets a few micrometers in size which 
have a porosity of around 50$\%$ \cite{simon}.
On the mesoscopic scale a single platelet is a collection of the stacked PbO layers \cite{lec}
(Fig.~\ref{fig:fig1}). It is expected that a layered structure 
will result in anisotropy in the transport properties and inter-platelets and/or interlayer charge transfer will 
limit the overall charge mobility.

Another common reason for the low charge drift mobility in 
polycrystalline compounds that must be considered in PbO
is charge carrier trapping.
Traps capture the photogenerated carriers or act as the 
scattering or recombination centers \cite{kabir}. 
Generally, the effect of the carrier trapping on the transport properties depends 
on the trap concentration.
The process of carrier de-trapping
from shallow traps is fast in comparison to 
the deep traps which tend to hold the carriers 
longer, thereby significantly impairing the temporal characteristics of the material. 
For x-ray application, the worst situation is 
when the carriers de-trapping becomes longer
than the collection time of the X-ray signal and as
such the de-trapped carriers contribute to the appearance 
of image artifacts known as 'ghosting' \cite{rowland}. This limits the application of the
PbO-based detectors in real-time imaging procedures.

Thermally deposited PbO has an oxygen deficiency \cite{chienes,bigelow,scanlon}.
Although the effect of the O vacancies on transport in PbO is not clearly understood, 
it was found that thermal annealing in pure oxygen increases the electrical conductivity
that can be attributed to the reduction of the oxygen vacancy concentration \cite{chienes}.
Unfortunately, high-temperature annealing is not practical for PbO layers deposited over 
an imaging matrix. Therefore, comprehensive studies of PbO structure and defect formation 
in thermally evaporated PbO layers are needed to improve the temporal characteristics of the PbO compound.
A growth process must be developed to reduce the trap concentration and ultimately to improve the
performance of PbO for real time x-ray detector applications. 
We started our investigations by modelling $\alpha$-PbO crystal lattice, revealing the crystal 
structure parameters and verifying them with available theoretical 
and experimental data. Due to limitation of
the software used, the dispersive interactions have been neglected in our studies. 
The effect of this assumption on results is discussed in Sec. II.
Further, the vacancies have been induced
into the lattice of $\alpha$-PbO and 
defect formation energies, energetic location of the defects within the band gap 
and potential of their transition 
to the charged states have been investigated.
In order to show the consistency 
of our results with the experimental data, we completed our work 
with measurements of the dark current on the ITO/PbO/Au samples
with an analysis of the trap participation in the 
charge transport process.

\section{Calculation technique and experiment}
\subsection{Calculation technique}
In our study we applied the density functional theory (DFT) available in
the Wien2k package \cite{wien} which utilizes the 
full-potential augmented plane-wave method.
All calculations of the electronic structure were performed with the
Perdew-Burke-Ernzerhof parametrization \cite{GGA} of the generalized gradient approximation (GGA) to DFT.

Calculations of the formation energy of defects (the cohesive properties) with DFT methods
require attention to many parameters and calculations with high precision have to be implemented.
Applying the appropriate energy cut off to separate the core and valence states 
we treated $5p$, $5d$, $6s$ and $6p$ electrons of Pb atoms 
and $2s$ and $2p$ electrons of O atoms as valence electrons.
The inclusion of the $5d$ states of Pb atom in the valence states 
(not in the core states) is required to 
introduce a proper description of bonding and orbital hybridization of Pb atoms with O atoms.
Indeed, the orbital energy of the $5d$ electrons in Pb atoms
and $2p$ electrons in O atom
are located in the same energy range \cite{terp}.
There is one more 
parameter needing special attention when the cohesive properties 
are studied, the so-called $RK_{max}$ parameter,
which is the product of the atomic sphere radius and the plane-wave cutoff 
in $k$-space. In our calculation it was assigned as 8. 
The Brillouin zone of a primitive cell for most of the calculations 
was set to a 11$\times$11$\times$8 Monkhorst-Pack mesh. 
When the supercell procedure was used, the size of the Monkhorst-Pack mesh was adjusted 
to the size of the supercell, i.e. the mesh was appropriately reduced 
as the supercell was enlarged. The calculations of the 
formation energy of the native point defects 
have been done with the 
sufficiently large supercell of 108-atom size 
(3$\times$3$\times$3 array of the primitive unit cells) and 4$\times$4$\times$3 Monkhorst-Pack mesh.

The optimization procedure for the $\alpha$-PbO lattice was performed based on minimization of the forces \cite{forces} 
and it provided us with the following lattice parameters: lattice constants $a_0=4.06$ \AA\ (within the layer) and $c_0=5.51$ \AA\ 
(inter-layer distance) with the ratio $c_0 / a_0=1.357$ and 
the Pb-O bond length of 2.35 \AA. The achieved magnitude for $a_0$ is identical to the value 
obtained with GGA in Ref. \cite{venk} and is in excellent agreement 
with the experimental value of 3.96 \AA obtained in Ref. \cite{lec}. 
The calculated length of Pb-O bond is in agreement with other 
theoretical studies \cite{walsh} and correlates well with an experimental value of 2.32 \AA\ \cite{lec}. 
The second lattice parameter $c_0$ agrees very well with GGA calculations performed in 
Ref.\cite{walsh, Oleg} but is larger than the experimentally determined value of 5.07 \AA\ \cite{venk}.

The mismatch in lattice parameter $c_0$ occurs due to limitations of GGA functional 
which does not include into account the dispersive interactions. 
In attempt to compensate for this limitation, we have used the experimental 
value of the lattice parameter $c_0$=5.07 \AA\ \cite{venk} that induces a 
reduction in layer separation. 
As a result, an increase in the interlayer interaction 
strength has contributed into a raise of the layers binding energy from 0.013 eV/atom to 0.016 eV/atom
being surprisingly small (calculated as the difference in the total energy between 
the systems of a single layer and two layers of $\alpha$-PbO). 
However, it has caused significant suppression in the band gap size. 
If for the optimized lattice constant $c_0=5.51$ \AA\ the 
indirect gap $\boldsymbol{\Gamma}-\mathbf{M}^*$ is 1.8 eV \cite{mine} 
which is in good agreement with experimentally found optical gap of 1.9$\pm$0.1 eV \cite{thang},
with implementation of the lattice parameter $c_0$=5.07 \AA\ \cite{venk}, the gap shrinks by 0.22 eV. 
The important distinction of the $\alpha$-PbO crystal structure is that the layers are held together 
by the weak orbital overlap of the $6s^2$ lone pairs \cite{terp} while reduction of $c_0$ leads 
to an overestimation of the interlayer overlap of these orbitals by GGA. 
Since the indirect gap is defined by strength of the interlayer interactions, 
it decreases with $c_0$ suppression \cite{mine}. 
Similar was observed for SnO which has the same lattice type as $\alpha$-PbO
(129P4/nmm space group) for which inclusion of the dispersive interactions
while correcting the lattice parameters 
caused unreasonable reduction in the band gap sizes $E_G$ \cite{allen}.
However, the formation energies of vacancies did not show strong dependence 
on the lattice parameter $c_0$. The vacancy states are localized entirely 
within the single layer, such as their formation energy is affected strongly 
only by the lattice constant $a_0$ and length of the Pb-O bond, 
but contribution of the interlayer interactions defined by $c_0$ is negligibly small. 
For the purposes of this work, the large discrepancy in the band gap size makes 
it difficult to define correctly an appearance of defects inside the band gap. 
Therefore, in order to achieve the meaningful results on both, location of the defect states 
inside the band gap and their formation energy, we found more justified to 
use the optimized lattice constant $c_0=5.51$ \AA\ (see Sec. III). 

The value of the interlayer interactions in order of 0.013 eV/atom 
(0.016 eV/atom for $c_0$=5.07 \AA) is lower than thermal energy at room temperature 
$kT\simeq 0.026$ eV and much lower than in most other solid state materials. For example,
in graphite which consists of stacked graphene layers, 
the interlayer interactions were found to be 0.052 eV/atom \cite{graphite}.
The extremely low magnitude of interactions in $\alpha$-PbO is similar to the inter-molecular interactions
in $\pi$-conjugate organic systems \cite{DNA}. For $\pi$-conjugate organic systems the low interlayer interaction
is the primary reason for the extremely low carrier mobility. 
Since in PbO the interlayer interaction is half of $kT$ at room temperature, 
we anticipate that the interlayer electron transport in $\alpha$-PbO
by all potential transport mechanisms would be insufficient, meaning 
that electron mobility in this direction is extremely low.
This statement is supported by the almost dispersionless 
valence bands observed in the band diagram of $\alpha$-PbO \cite{terp}.

\subsection{Experimental details}
Thick (40 $\mu$m) PbO layers were deposited on ITO-covered 
Corning glass substrates by thermal evaporation of the PbO powder 
(purity 99.9999 $\%$) in a vacuum of 0.2 Pa under additional molecular oxygen flow.
The growth rate was 1 $\mu$m/min. The substrate was kept at 120$^{\circ}$ C to suppress the
growth of $\beta$-PbO \cite{simon} (grown PbO layers may contain a trace 
of other lead-compounds) and to achieve good adhesion to the substrate.
Subsequently, gold contacts were deposited through the contact
mask by a spattering technique. The resulting ITO/PbO/Au
structures were biased to different electric fields
(ITO was negatively biased)
and time dependence of the dark current density was measured
automatically every second for 50 minutes.

\section {Results and discussion}
\subsection{Native point defects}
As it was mentioned above, in practice thermally grown polycrystalline $\alpha$-PbO 
contains large amount of the O vacancies \cite{chienes,bigelow,scanlon}.
Therefore, in this work we consider the formation of the native point defects 
(i.e. O or Pb vacancies) assuming that no impurities
responsible for formation of other type of defects
are present. This assumption is quite reasonable when 
PbO is grown in vacuum employing only molecular oxygen flow. We applied a supercell 
approach in which the larger is the supercell size, the smaller is the interaction between the 
repeated units (neighbouring supercells each containing a defect). A truly isolated defect 
should show the dispersion-less flat band on the band diagram
which was indeed received with a supercell of 108-atom size (3$\times$3$\times$3 array).

\begin{figure}
\includegraphics[scale=0.6]{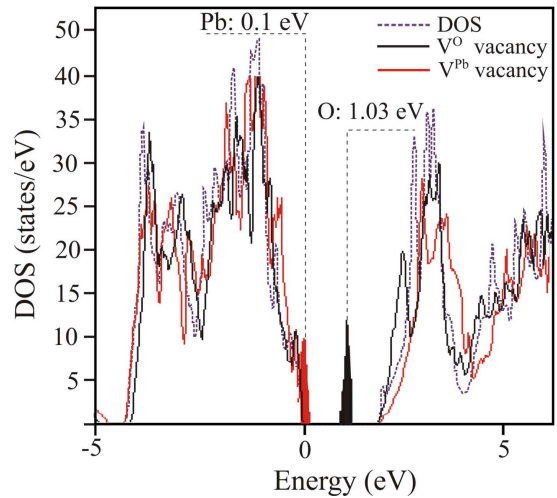}
\caption{\label{fig:fig2} Colour on-line. Density of states calculated for 
defect free $\alpha$-PbO (dashed line) and the same system with defects (solid lines): DOS for  
the O vacancy is shown by red solid line 
while for the Pb vacancy it is marked by a black line. The energetic location of the 
vacancy states inside the band gap relatively the top of the valence band is 
$E_{D}$($V^{\operatorname{O}}$)=1.03 eV and $E_{D}$($V^{\operatorname{Pb}}$)=0.01 eV 
for the O vacancy and, the Pb vacancy, respectively.}
\end{figure}

We examined the formation of vacancy defect by removing one of the
corresponding atoms from the $\alpha$-PbO lattice and then optimized
its geometry with respect to the internal degrees of freedom.
As a single defect is induced, the lattice rearrangement around the defect 
site occurs. Within the $\alpha$-PbO lattice, the layer of O atoms are 
tightly sandwiched between Pb atoms (see Fig.~\ref{fig:fig1}), i.e. the Pb atoms are located on the side 
of the layers holding the skeleton of each layer. 
Therefore, removal of the 
Pb atom from the lattice induces a significant lattice rearrangement,
while in the case the O atom of small atomic radius is removed, 
the distortion of the lattice is minimal.
Thus, removal of Pb atom ($V^{\operatorname{Pb}}$) initiates the enlargement of 
the distance between O atoms which were bonded to
Pb site, and each O atom moves apart from the defect site by 0.22 \AA\ 
due to a repulsion felt by the O ions. 
In contrast, the lattice modification induced by the O vacancy is almost unnoticeable: 
Pb atoms move only by 0.07 \AA\ toward the vacancy site.

We show in Fig.~\ref{fig:fig2} an alteration in the density of states (DOS)
as defects are induced into the $\alpha$-PbO lattice.
Insertion of either O or Pb vacancy creates the defect level
inside the band gap and the electronic density for the defect state 
is strongly localized in both cases. The electron density distribution outside the band gap
is not significantly affected by the presence of
defects since only O:2p$^4$ and Pb:6p$^2$ electrons 
participate in formation of the Pb-O bond (these states are located close to the band gap edges).
Our results indicate that O vacancy ($V^{\operatorname{O}}$) 
forms the defect level near the midgap.
The energetic location of this defect level is $E_{D}$($V^{\operatorname{O}}$)=1.03 eV 
above the top of the valence band $E_V$ as shown in Fig.~\ref{fig:fig2}. 
The O vacancy in its uncharged 
state is already filled with two electrons. 
Because O atom forms 4 bonds with nearest Pb atoms in the $\alpha$-PbO lattice, 
a removal of a single O atom leaves 0.5 unbounded electrons on 
each Pb atom that overall results in occupation of this defect level by two electrons. 
The Pb vacancy ($V^{\operatorname{Pb}}$) induces the defect energetically located
close to the top of the valence band 
with $E_{D}$($V^{\operatorname{Pb}}$)=0.1 eV and this vacancy is not filled with electrons.

\subsection{The formation energy of the vacancies}
The formation energy of a vacancy is an important parameter as it determines 
how likely a vacancy will be generated in the compound under given
growth conditions.
The formation energy is mainly defined by several parameters: 
($i$) type of the crystal structure as energy required to 
remove an atom from the crystal structure depends on the strength of
the electronic interactions within the lattice, ($ii$) the final state of the 
removed species, ($iii$) the characteristics of the environment, i.e. 
growth conditions. 

The formation energy of a defect $D$ in charge state $q$ can be defined as \cite{walle}:
\begin{equation}
\Delta E^f(D)=E_{tot}(D^q)-E_{tot}(S)+\sum_{i}n_{i}\mu_{i}+q(E_F+E_V+\Delta V)
\label{eq:one}
\end{equation}
where $E_{tot}(D^q)$ and $E_{tot}(S)$ 
are the total energy of the system containing the single defect and 
defect-free system, respectively. 
$n_{i}$ indicates a number of $i$-atoms removed while
$\mu_{i}$ is the chemical potential of those atoms.
($E_F+E_V$) is the position of the Fermi level relative to
the valence band maximum ($E_V$). $q$ defines the charge of the state (+2/+1/0/-1/-2). 
For the charged point defects, the position of the valence band $E_V$
has to be corrected with $\Delta V$ calculated through alignment of the reference potential 
in defective supercell with that in bulk $\alpha$-PbO (for details see Ref. \cite{walle}).
The formation energy of the defects has been also corrected 
with so-called band gap error $\Delta E_G$ \cite{togo} which is 
defined as a difference between the direct band gap 
$\boldsymbol{\Gamma}$-$\boldsymbol{\Gamma}^*$=1.94 eV
and the experimental optical band gap of 1.9 eV \cite{thang}. 
In this particular case, the contribution of $\Delta E_G$=-0.04 eV into the formation energy is minor because of 
good agreement of the theoretical value and experimental data
(a value of the corrected band gap of suprecell \cite{togo} is 1.89 eV).
The formation energy has been increased by this band gap correction 
$m \Delta E_G$, where $m$ is the number of the electron at the defect site.

The chemical potentials are defined as following $\mu_{i}=E_{tot}(i)+\mu_{i}^*$, 
where $\mu_{i}^*=\mu_{i}^0+kT\cdot ln(p/p^0)$ is the part related to 
real growth conditions: the partial pressure $p$ and temperature $T$ 
($\mu_{i}^0$ is an alteration to the chemical potential 
induced by change of temperature from 0 to T under the standard pressure $p^0$).  
However, not willing to speculate on the Pb and O partial pressures 
used during deposition, we consider the extreme cases, i.e. the Pb-rich or O-rich growth conditions 
($\mu_{\operatorname{(Pb)}}$=$\mu_{\operatorname{(Pb)[bulk]}}$ and $\mu_{\operatorname{(O)[O_2]}}$, respectively)
as it was suggested in Refs. \cite{walle,togo,allen,zheng}.
The chemical potentials for the extreme cases can be evaluated through the 
standard enthalpy of formation $\Delta_f H^0(\operatorname{PbO})$ as \cite{walle}: 
\begin{equation}
E_{tot}(\operatorname{PbO})=\mu_{\operatorname{(Pb)[bulk]}}+\mu_{\operatorname{(O)[O_2]}}+\Delta_f H^0(\operatorname{PbO})
\label{eq:rate1}
\end{equation}
where $E_{tot}(\operatorname{PbO})$ is the 
total energy of the product; $\mu_{\operatorname{(Pb)[bulk]}}$ and $\mu_{\operatorname{(O)[O_2]}}$
are the chemical potentials of bulk Pb and O$_2$ molecule, respectively.
Therefore, $\Delta_f H^0(\operatorname{PbO})$ is an important parameter in definition 
of the chemical potentials. In the calculation of $\Delta_f H^0$ for oxides
the main discrepancy between the theoretical and experimental data is known to come from the 
binding energy of the O$_2$ molecule ($\Delta_f H^0$(O$_2$)) used in 
definition of the chemical potential
$\mu_{\operatorname{(O)}[O_2]}=\frac{1}{2}(2E_{tot}(\operatorname{O})+\Delta_f H^0$(O$_2$)+$\mu_{(\operatorname{O}_2)}^*)$ 
\cite{wang,zheng,hammer}.
To disregard this error in our calculations 
we used the experimental value of $\Delta_f H^0$(O$_2$)=-5.23 eV \cite{exp} 
(best theoretical estimation is $\Delta_f H^0$(O$_2$)=-6.01 eV \cite{wang}).
With that assumption we obtained $\Delta_f H^0$(PbO)=-2.92 eV per Pb-O pair 
which is in appropriate agreement with the
experimental value of $\Delta_f H^0$(crystal $\alpha$-PbO)=-2.29 eV \cite{JANAF}.

\begin{figure}
\includegraphics[scale=0.5]{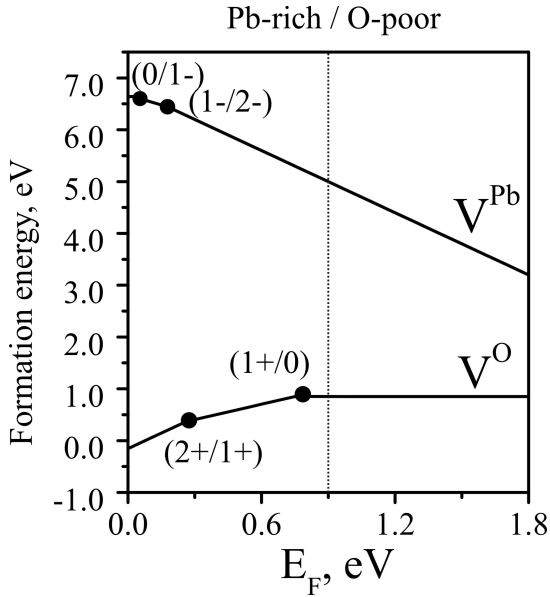}
\caption{\label{fig:fig3} The formation energy of 
the defects $\Delta E^f(D)$ for $V^{\operatorname{O}}$ and $V^{\operatorname{Pb}}$ 
for Pb-rich/O-poor limit. The charge states for which added electron or hole remains localized on the vacancy site 
are shown (1+/2+ states for the $V^{\operatorname{O}}$ vacancy and 
2$-$/1$-$ states for the $V^{\operatorname{Pb}}$ vacancy).}
\end{figure}

It is known that regardless the deposition techniques used, the PbO layers 
are not stoichiometric and has deficit of oxygen
\cite{chienes,bigelow,scanlon}. Hence, we consider the Pb-rich/O-poor conditions
for which the O-poor limit has been assigned to
$\mu_{\operatorname{(O)}}^* =\Delta_f H^0(\operatorname{PbO})$ \cite{togo} 
($\mu_{\operatorname{(O)}}^*$=-2.92 eV), while Pb-rich limit has been found 
from relation 
$\Delta_f H^0(\operatorname{PbO})=\mu_{\operatorname{(Pb)}}^*+\mu_{\operatorname{(O)}}^*$ ($\mu_{\operatorname{Pb}}^*$=0 eV). 
The formation energy of 
the defects $\Delta E^f(D)$ for $V^{\operatorname{O}}$ and $V^{\operatorname{Pb}}$ for the different charge states 
calculated with help of Eq. \ref{eq:one} are presented in Fig.~\ref{fig:fig3}.

\begin{figure}
\includegraphics[scale=0.50]{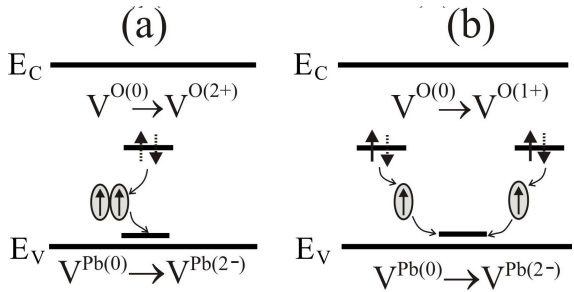}
\caption{\label{fig:fig4} Schemes showing the mechanisms of ionization 
of the neutral vacancies through the electron exchange
($V^{\operatorname{Pb}}$ accepts two or single electron occupying the 
$V^{\operatorname{O}}$ vacancy).}
\end{figure}

The O vacancy in its neutral state is occupied by two electrons.
If the O vacancy drops electron (1+ charged states), 
its formation energy is reduced.
The Pb vacancy in its uncharged state is empty $V^{\operatorname{Pb(0)}}$ 
and its formation energy is comparably high, but is lowered
if vacancy accepts electrons (1$-$/2$-$ charged states).
Therefore, both the O and Pb vacancies ($V^{\operatorname{O}}$ and $V^{\operatorname{Pb}}$) 
intend to appear in the opposite charged states. 
The neutral O vacancy would prefer to give away one electron 
to reduce its formation energy. To conserve the electroneutrality of material 
there are only the Pb vacancies that can accept 
electrons under the equilibrium conditions. 
This makes the Pb vacancy a compensation center for the O vacancy. 
The considered mechanisms of the charge exchange between vacancies are presented 
Fig.~\ref{fig:fig4} (a) and (b). In case of thermodynamic equilibrium,
when formation energies for both types of vacancies are equal and, therefore, 
their concentrations are equal as well,
vacancies would become doubly ionized ($V^{\operatorname{Pb(2-)}}$ and $V^{\operatorname{O(2+)}}$) 
utilizing the mechanism of the electron exchange as presented in Fig.~\ref{fig:fig4} (a).

However, for the Pb-rich/O-poor limit considered in present work 
the formation energies of the vacancies are
different from the thermodynamic equilibrium (see Fig.~\ref{fig:fig3}).
As seen in Fig.~\ref{fig:fig3}, the
formation energy of the O vacancy is much lower than the Pb vacancy. 
This presumes much higher concentration of the O vacancies so that 
they will be only partially compensated by the Pb vacancies.
Based on achieved magnitude of the formation energies 
we expect the Pb vacancies to appear preferably
in their (2$-$) states while the O vacancies 
to be formed in the different charged states, (0)/(1+).
In this case, ionization of the 
Pb vacancy to the $V^{\operatorname{Pb(2-)}}$ state can occur with participation 
of two O vacancies such as each $V^{\operatorname{O(0)}}$ donates single electron 
becoming ionized only to the $V^{\operatorname{O(1+)}}$ state as presented in Fig.~\ref{fig:fig4} (b).
The O vacancies in the $V^{\operatorname{O(1+)}}$ state are still occupied
by one electron, and the larger the difference in O/Pb vacancy concentration, 
the larger amount of the non-compensated O vacancies.
This behaviour helps in understanding of
the experimentally observed $n$-conductivity of PbO \cite{chienes,bigelow,scanlon}. 
Both neutral and singly charged O vacancies 
($V^{\operatorname{O(0)}}$ and $V^{\operatorname{O(1+)}}$) act as 
$n$-type donor. Moreover, in this case, 
a pinning of the Fermi level position slightly above
the midgap (0.95 eV below the conduction band)
observed experimentally \cite{broek} can be assigned
to $n$-type doping induced by the O vacancies.
Previously, the pinning was associated with the surface states 
at the crystallites boundaries but nature of those states was unknown \cite{broek}.
A remarkable agreement between the Fermi level position predicted in
Ref.\cite{broek} and position of the O vacancy states 
found here (see Fig.~\ref{fig:fig2}) suggests that the Fermi level is stabilized by the 
presence of the O vacancies.

Therefore, we anticipate that the O vacancies would affect the 
transport and photogeneration in lead oxide more significantly than the Pb vacancies. 
Indeed, shallow traps for holes created by the ionized Pb vacancies 
might slightly reduce the hole mobility which is already low due to the extremely heavy holes \cite{mine}, 
but much deeper O vacancies when they are ionized
(see Fig.~\ref{fig:fig2}) would not only
slow down the electron propagation in the conduction band through trapping, but can additionally act 
as the recombination centers. Therefore, because a contribution of the deep traps in the charge transport
is known to impair significantly the current decay \cite{kabir1,street}, 
the temporal behaviour of the dark conductivity can
be used to confirm a presence of the O vacancies.

\subsection{Dark current kinetics}
\begin{figure}
\includegraphics[scale=0.8]{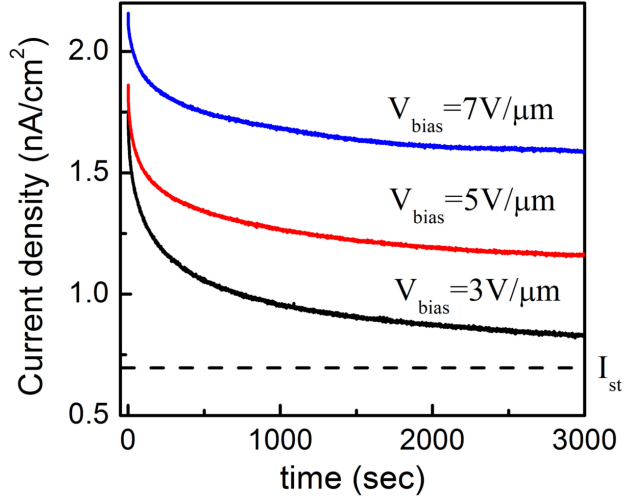}
\caption{\label{fig:fig5} Colour on-line. The time dependence of the dark current density
for the different bias applied. $I_{st}$ is the steady state current
which is reached at 1.5$\times 10^4$ sec.}
\end{figure}

The dark current kinetics is a sensitive measure of
the electronic properties of a material and is used here to
describe the effect of point defects on the conductivity in PbO layers.  
The results of time dependence of the dark current density 
for selected biases are shown in Fig.~\ref{fig:fig5}.
As it is seen from Fig.~\ref{fig:fig5}, after bias voltages are
applied dark current decays slowly reaching a steady state 
value after about 250 minutes. The steady-state current density
depends on electric field and increases by a factor of 2 when bias is increased from 3 to 7 V$\mu$m. 
Similar behaviour of the dark current was observed by Mahmood and Kabir
in amorphous selenium (a-Se) multilayer $n-i-p$ structures \cite{kabir1} 
and by Street in hydrogenated amorphous silicon (a-Si:H) $p-i-n$ structures \cite{street, street3}. 
Mahmood and Kabir explain dark current decay in a-Se multilayer structure by carrier trapping 
within comparatively thick (few $\mu$m \cite{rev}) $n$- and $p$- layers which 
induces screening of the electric field at 
the metal/$n$- or $p$-layer interfaces.
The subsequent redistribution of the electric field suppresses
carrier injection from metal contacts and reduces the dark current which is mainly controlled by the injection.

An alternative model is suggested by Street who explains similar
dark current kinetics by depletion of charge from the bulk of $i$-layer
assuming that dark current is controlled by thermal generation from 
defects states in the gap. As traps are depleting, the quasi Fermi 
level moves toward the midgap and the thermal generation current decreases \cite{street}.
As PbO samples studied here are uniform, we have to assume 
that the electric field is also uniform across the layer
(neglecting thin pre-contact areas). Therefore, the model developed by Street
is more applicable in our case. Neglecting charge carrier injection under the
applied low biases, we can assume that the thermal generation current arises
from the excitation from the O vacancies occupied by electrons to the conduction band.
Hence the dark current decay can be described by the time dependent quasi
Fermi level position as is was suggested in Ref. \cite{street}. 
Without speculating on the degree of compensation in our layers
we assume that quasi Fermi level is initially located at 1.03 eV
from the valence band, i.e. the activation energy is 0.77 eV. 
After an electric field is applied,
the occupancy of the O vacancies changes as electrons are
emitted to the conduction band and the quasi Fermi level shifts
toward the midgap. Once traps are fully depleted quasi Fermi level
approaches the equilibrium Fermi level and the thermal generated
current saturates at electric field dependent steady-state value. 

Although at that point we provide just qualitative analysis of
dark current kinetics in PbO layers, it allows us to link defects
(namely, the O vacancies) with transport properties in this material.
It has to be mentioned that steady-state dark current is
extremely low (and much lower than in a-Se \cite{kabir1})
that is very encouraging for the application of PbO layers
in direct conversion medical x-ray imaging detectors \cite{detectors}.

\section{Conclusion}
First-principles density-functional calculations
were used to calculate the electronic properties of polycrystalline
$\alpha$-PbO and the formation of native point defects
(namely, O and Pb vacancies) in this material.
It was found that the O vacancies induce very deep
donor level close to the midgap at 1.03 eV above the valence band.
In contrast, the Pb vacancies create shallow defect level at
just 0.1 eV above the valence band which acts as acceptor.
Under applied bias the ionized O vacancies in PbO would act as the deep traps and 
recombination centers for the electrons in the conduction band, 
while the Pb vacancies are the shallow traps for holes in the valence band.

The formation energies
of the defects in their neutral charge states are comparatively small:
0.85 eV for the O vacancy and 6.64 eV for the Pb vacancy ($E_F$ is 
assigned to the midgap and the Pb-rich/O-poor growth conditions are considered) 
and are further reduced if a vacancy appears in its energetically
favourable charged state. For example, for the doubly ionized (2-)
Pb vacancy the formation energy is reduced to 4.99 eV.
The electron exchange between vacancies initiates ionization of the vacancies,
but for Pb-rich/O-poor growth conditions when concentration of the vacancies 
is not balanced, most of the O vacancies remains unionized, i.e. in
their (0) uncharged state or (1+) charged state in which
the O vacancies are occupied with electrons.

The presence of defects and differential trapping of electrons and
holes were predicted by others to explain space charge limited
photoconductivity in PbO \cite{hughes} and to model x-ray sensitivity,
modulation transfer function (MTF) and detective quantum efficiency (DQE)
of the PbO x-ray detector \cite{kabir}. 
The results presented here agree well with these studies and
provide insight into the nature of defects in PbO clarifying
their electronic and charge states and explaining why vacancies
exist in high concentration in thermally evaporated PbO layers.
Moreover, our own experimental results on time dependence of
the dark current density suggest that this is the field dependent
occupancy of O vacancies that governs the dark current kinetics.
Thus, the O vacancies are occupied with electrons
and because these centers are located close to 
the midgap of PbO, a process of detrapping of the vacancies is 
slow thus impairing the temporal characteristics of compound.
 
Since O vacancies play more essential role in the transport 
properties of PbO layers, material science solutions must be
found to improve PbO layers deposition techniques in order to suppress
their appearance. Methods to consider include thermal evaporation
with optional low energy O ion bombardment or passivation
of vacancies by post-growth annealing in oxygen atmosphere.

\section*{Acknowledgement}
Authors are thankful to Dr. Matthias Simon (X-ray Imaging Systems, Philips Research) 
for numerous stimulating discussions and Giovanni DeCrescenzo for technical support 
in conducting the dark-current measurements. 
Financial support of Ontario Research Fund- Research Excellence program is highly acknowledged.


\begin{thebibliography}{99}
\bibitem[\ddag]{byline} Electronic mail: berashej@tbh.net
\bibitem{simon}
Simon M, Ford R A, Franklin A R, Grabowski S P, Menser B,
Much G, Nascetti A, Overdick M, Powell M J and
Wiechert D U 2005 {\it IEEE Transactions on Nuclear Science} {\bf 52}, 2035.
\bibitem{Sasha}
Kasap S, Frey J B, Belev G, Tousignant O, Mani H, Greenspan J, Laperriere L,
Bubon O, Reznik A, DeCrescenzo G, Karim  K S and Rowlands J A 2011
{\it Sensors} {\bf 11} 5112.
\bibitem{hughes}
Hughes R C and Sokel R J 1981 {\it J. Appl. Phys.} {\bf 52} 6743.
\bibitem{lec}
Leciejewicz J 1961 {\it Acta Cryst.} {\bf 14} 1304.
\bibitem{kabir}
Kabir M Z 2008 {\it J. Appl. Phys.} {\bf 104} 074506.
\bibitem{rowland}
Rau A W, Bakueva L and Rowlands J A 2005 {\it Med. Phys.} {\bf 32} 3160.
\bibitem{chienes}
Hwang O, Kim S, Suh J, Cho S and Kim K 2011 
{\it Nuclear Instruments and Methods in Physics Research A} {\bf 633} S69.
\bibitem{bigelow}
Bigelow J E and Haq K E 1962 {\it J. Appl. Phys.} {\bf 33} 2980.
\bibitem{scanlon}
Scanlon D O, Kehoe A B, Watson G W, Jones M O, David W I F, 
Payne D J, Egdell R G, Edwards P P and Walsh A 2011 {\it Phys. Rev. Lett.} {\bf 107} 246402.
\bibitem{wien}
Blaha P, Schwarz K, Madsen G K H, Kvasnicka D and Luitz J {\it Wien2k: An Augmented 
Plane Wave + Local Orbitals Program for Calculating Crystal Properties: 
Karlheinz Schwarz}, (Techn. Universit\"at Wien, Austria, 2001)
\bibitem{GGA}
Perdew J P, Burke K and Ernzernof M 1996 {\it Phys. Rev. Lett.} {\bf 77} 3865.
\bibitem{terp}
Terpstra H J, de Groot R A and Haas C 1995 {\it Phys. Rev. B} {\bf 52} 11690.
\bibitem{forces}
Yu R, Singh D and Krakauer H 1991 {\it Phys. Rev. B} {\bf 43} 6411.
\bibitem{venk}
Venkataraj S, Geurts J, Weis H, Kappertz O, Njoroge W K,
Jayavel R and Wuttig M 2001 {\it J. Vac. Sci. Technol. A} {\bf 19} 2870.
\bibitem{walsh}
Walsh A and Watson G W 2005 {\it J. Solid State Chemistry} {\bf 178} 1422.
\bibitem{Oleg}
Rubel O and Potvin A 2011 {\it AIP Conf. Proc.} {\bf 1368} 85.
\bibitem{mine}
Berashevich J, Semeniuk O, Rowlands J A and Reznik A 2012 {\it EPL} {\bf 99} 47005.
\bibitem{thang}
Thangaraju B and Kaliannann P 2000 {\it Semicond. Sci. Technol.} {\bf 15} 542.
\bibitem{allen}
Allen J P, Scanlon D O, Parker S C and Watson G W 2011 {\it J. Phys. Chem. C} {\bf 115} 19916.
\bibitem{graphite}
Zacharia R, Ulbricht H, Hertel T 2004 {\it Phys. Rev. B} {\bf 69} 155406.
\bibitem{DNA}
Berashevich J and Chakraborty T 2008 {\it J. Chem. Phys.} {\bf 128} 235101.
\bibitem{togo}
Togo A, Oba F, Tanaka I and Tatsumi K 2006 {\it Phys. Rev. B.} {\bf 74} 195128.
\bibitem{JANAF}
NIST-JANAF Thermochemical Tables. http://www.kinetics.nist.gov/janaf
\bibitem{walle}
Van der Walle C G and Neugebauer J 2004 {\it J. Appl. Phys.} {\bf 95} 3851.
\bibitem{wang}
Wang L, Maxish T and Ceder G 2006 {\it Phys. Rev. B.} {\bf 73} 195107.
\bibitem{zheng}
Zheng J X, Ceder G, Maxisch T, Chim W K and Choi W K 2007
{\it Phys. Rev. B.} {\bf 75} 104112.
\bibitem{hammer}
Hammer B, Hansen L B and Norskov J K 1999 {\it Phys. Rev. B} {\bf 59} 7413.
\bibitem{exp}
Pople J A, Gordon M H, Fox D J, Raghavachari K and Curtiss L A 1989 {\it J. Chem. Phys.} {\bf 90} 5622.
\bibitem{japan}
Wasa K and Hayakawa S 1969 {\it Jap. J. Appl. Phys.} {\bf 8} 276.
\bibitem{broek}
Wolfe W L {\it Optical physics and engineering}, {Plenum Press, New York, London, 1971}; 
van der Broek J 1967 {\it Philips Res. Rep.} {\bf 22} 367.
\bibitem{kabir1}
Mahmood S A and Kabir M Z 2011 {\it J. Vac. Sci. Technol. A} {\bf 29} 031603.
\bibitem{street}
Street R A 1990 {\it Appl. Phys. Lett.} {\bf 57} 1334.
\bibitem{street3}
Street R A, Ready S E, Lemmi F, Shah K S, Bennett P and Dmitriyev Y 1999 {\it J. Apll. Phys.} {\bf 86} 5.
\bibitem{rev}
Kasap S, Frey J B, Belev G, Tousignant O, Mani H, Laperriere L, Reznik A and Rowlands J A 2009 
{\it Physica Status Solidi (b)} {\bf 246} 1794.
\bibitem{detectors}
Kabir M Z, Kasap S O and J. A. Rowlands J A, 
{\it in Springer Handbook of Electronic and Photonic Materials}, 
edited by S. O. Kasap and PeterCapper (Springer, Heidelberg, Chap. 48, 2006).
\end{thebibliography}
\end{document}